\documentclass{cernyrep} 
\usepackage{epsfig}
\usepackage[T1]{fontenc}
\usepackage{graphicx}
\usepackage{epstopdf}
\usepackage{morefloats}
\usepackage{placeins}
                       
\usepackage{varwidth}
\usepackage{xcolor}
\usepackage[customcolors]{hf-tikz}

\usepackage{ifthen}
\usepackage{array,tabularx}
\usepackage{amssymb}
\usepackage{xcolor}     
\usepackage{lipsum}     
\usepackage[standard]{ntheorem}   
\usepackage{mdframed}   
\usepackage{enumitem}

\usepackage{tikz,pgfplots}
\pgfplotsset{compat=newest}
\usepgfplotslibrary{groupplots}
\usepgfplotslibrary{external} 

\theoremstyle{break}
\theoremheaderfont{\bfseries}
\newmdtheoremenv[%
linecolor=gray,leftmargin=40,%
rightmargin=40,
backgroundcolor=gray!40,%
innertopmargin=10pt,%
ntheorem]{myinterlude}{Interlude}[section]

\newmdtheoremenv[%
linecolor=gray,leftmargin=40,%
rightmargin=40,
backgroundcolor=red!30,%
innertopmargin= 5pt,%
ntheorem]{myinterlude2}{Warning}

\newmdtheoremenv[%
linecolor=gray,leftmargin= 5,%
rightmargin= 5,
backgroundcolor=blue!30,%
innertopmargin= 2pt,%
ntheorem]{myinterlude3}{}

\usepackage[small,bf,nooneline]{caption2}

\newcommand{\Myabstract}[1]{
\begin{quote}
{\large\bfseries{Abstract}}\\
\rule{14cm}{2pt}
\vskip 2mm
#1
\vskip-1mm
\rule{14cm}{2pt}
\end{quote}}

\begin{document}
\section{pyTRAIN - a modern TRAIN implementation}
\pgfkeys{/pgf/number format/.cd,1000 sep={}}
\newcommand{\app}{\mathrm{\sim}}
\newcommand{\ut}[1]{\,\mathrm{#1}}
\noindent
{\it Michi Hostettler, Xavier Buffat, Tobias Persson, Tatiana Pieloni, Jorg Wenninger} \\
{{ 
CERN, Geneva, Switzerland\
}}

\Myabstract{
The TRAIN code, developed in 1995 as a post-processor for second-order transport maps from MAD, has been used extensively at the LEP and the LHC to study self-consistent closed orbits, tunes and chromaticities of bunch trains under the presence of beam-beam long-range (BBLR) and PACMAN effects..

This paper presents a modern re-implementation of the TRAIN concept in Python using well-known numeric libraries (numpy, scipy) and an optional link to MAD-X via cpymad. This greatly improves the usability, maintainability and extensibility of the code. New functionality includes the support for arbitrary particle types, an arbitrary number and distribution of beam-beam interaction points, and the extrapolation of the beam-beam induced closed-orbit effects to arbitrary points in the machine. The code is benchmarked against the classic TRAIN code, and simulation results are compared to observations from LHC physics operation.}

\section*{Introduction}
In a collider with two beams of many tightly spaces bunches (``bunch trains'') which share a common vacuum chamber at least in part of the machine, every bunch encounters multiple bunches of the counter-rotating beam in different points of the machine. These encounters can be at a separation (``long-range'') or head-on. Since each bunch encounters a different set of bunches in the other beam, this yields different closed orbits for each bunch. As the beam-beam force encountered by each bunch itself depends on the separation to (and hence, on the closed orbits of) all encountered bunches along the ring, a self-consistent treatment is required to resolve these effects.

The TRAIN code \cite{ID6077725_bib:origtrain} has been developed at LEP for simulation of self-consistent bunch-by-bunch orbits in the many bunch case under the presence of beam-beam effects. It has since been applied to other multi-bunch machines, in particular the LHC, where it predicted the clear advantage of the horizontal-vertical alternating crossing scheme due to self-compensation of the beam-beam long-range tune shift \cite{ID6077725_bib:wernerlhcxing}.

TRAIN does not implement any element transport code (except for beam-beam deflections). Instead, it takes second-order transport maps (``sector maps'') between the beam-beam interaction points and the machine geometry (``survey'') for both beams of the collider as inputs, which have to be generated by another accelerator physics code such as MAD-X \cite{ID6077725_bib:madx} or Xsuite \cite{ID6077725_bib:xsuite}. A second order map $\mathcal{M}$ is given by it's zeroth ($K_i$), first ($R_{ij}$), and second order ($T_{ijk}$) components as \cite{ID6077725_bib:werneretienne_nonlin}
\begin{align}
(\mathcal{M}x)_i &= K_i + R_{ij} x_j + T_{ijk} x_j x_k
\end{align}

Initially, for each of the two beams, TRAIN builds the second order one-turn map $\mathcal{O}$ without beam-beam interactions as the concatenation of all input maps, and performs an initial search for the closed orbit $z$ as its fixed point. This initial closed orbit is later used as the starting point for the iterative process of finding the bunch-by-bunch closed orbit under the presence of beam-beam effects. At this stage, TRAIN also establishes the beam-beam interaction schedule, determining which bunches encounter at which beam-beam interaction points considering the bunch filling scheme and the machine geometry.

The closed-orbit problem for all bunches under the presence of beam-beam effects is then solved iteratively:
\begin{enumerate}[beginpenalty=10000]
    \item For each bunch $b$ of beam 1, at each beam-beam interaction point $i$, insert a second order map $\mathcal{BB}(z_2[b',i]-z_1[b,i])$ describing the coherent beam-beam force \cite{ID6077725_bib:bassetti_erskine, ID6077725_bib:xavier_coherentbb} exercised by the bunch $b'$ of the counter-rotating beam encountered at this point at a distance given by difference in closed orbits $z_1$ and $z_2$ of the two bunches, assuming a Gaussian charge distribution for both bunches.
    \item For each bunch $b$ of beam 1, build the one-turn map $\mathcal{O}_{1}[b]$ (including the added beam-beam maps) and solve for the closed orbit $z_1[b]$.
    \item For each bunch $b$ of beam 2, at each beam-beam interaction point $i$, insert a second order beam-beam map $\mathcal{BB}(z_1[b',i]-z_2[b,i])$ for the encounter with bunch $b'$ in beam 1.
    \item For each bunch $b$ of beam 2, build the one-turn map $\mathcal{O}_{2}[b]$ and solve for the closed orbit $z_2[b]$.
    \item If convergence is reached, i.e.~the closed orbits $z_1[b]$, $z_2[b]$ of all bunches in both beams did not change significantly during the last iteration, the problem is solved. Otherwise, loop to step 1.
\end{enumerate}

In the nominal LHC physics case ($\app 2400$ proton bunches, $\app 1.6\cdot10^{11}$ protons per bunch, 4 interaction points with long-range beam-beam encounters), convergence at a $10^{-9}\ut{m}$ level in closed orbits is typically reached within 4-5 iterations. Once the closed-orbit problem is solved, the resulting second order maps for each bunch can be further analyzed to obtain bunch-by-bunch coherent tunes, chromaticities and optics functions ($\beta$, $D$).

\section*{The pyTRAIN code}
The classic TRAIN program was developed in FORTRAN-77 at LEP \cite{ID6077725_bib:origtrain}. While it has later been extended to the LHC case \cite{ID6077725_bib:hansgrote,ID6077725_bib:wernerlhcxing,ID6077725_bib:tatianatrain,ID6077725_bib:arekphd,ID6077725_bib:michiphd,ID6077725_bib:arianatrain}, its extensibility and applicability to new scenarios was limited and the usage was error-prone due to its legacy code.

To overcome these limitations, the pyTRAIN code \cite{ID6077725_bib:pytrain}  is a complete re-implementation of the TRAIN concept in python. It uses well tested and established numerical primitives from the numpy \cite{ID6077725_bib:numpy} and scipy \cite{ID6077725_bib:scipy} numeric libraries for matrix operations and the numeric evaluation of the beam-beam force (Faddeeva function). All operations are implemented in python code, with the exception of the concatenation routine for second-order maps, which is implemented in Cython \cite{ID6077725_bib:cython} and compiled to native machine code for efficiency.

The pyTRAIN code exposes a python API and data classes to prepare the inputs, call the solver, and post-process the outputs. In addition to the bunch-by-bunch closed orbits, coherent tunes and chromaticities, pyTRAIN also calculates the bunch-by-bunch Twiss parameters ($\beta$ functions and dispersions) under the presence of beam-beam effects from the second-order maps.

To treat external systems correcting part of the beam-beam effects (e.g.~an orbit feedback system correcting for the average beam-beam kicks) in a self-consistent way, hooks are provided within the solver's iteration loop to call custom code after establishing the beam-beam maps (steps 1 and 3). This code can add or modify bunch-by-bunch second order maps prior to finding the closed orbits (steps 2 and 4)

\section*{Application to LHC Data}
The pyTRAIN code was applied to predict the impact of beam-beam effects in the LHC during proton physics operation in 2023 and 2024. The two beams of LHC are typically filled with $\app 2400$ bunches spaced by $25\ut{ns}$ each during proton physics operation. For most part of the $27\ut{km}$ ring, they circulate in different vacuum chambers. In the inner part of the straight sections $165\ut{m}$ around each of the 4 interaction points (IPs), the beams enter a common vacuum chamber, resulting in up to 45 parasitic long-range beam-beam encounters per IP side, in addition to the 4 head-on (or, in case of offset levelling, partially separated) encounters at each of the 4 IPs \cite{ID6077725_bib:lhcdesign,ID6077725_bib:prablumiscan}.

\subsection*{Benchmarking and Validation}
\begin{figure}[h]
\begin{center}
\includegraphics{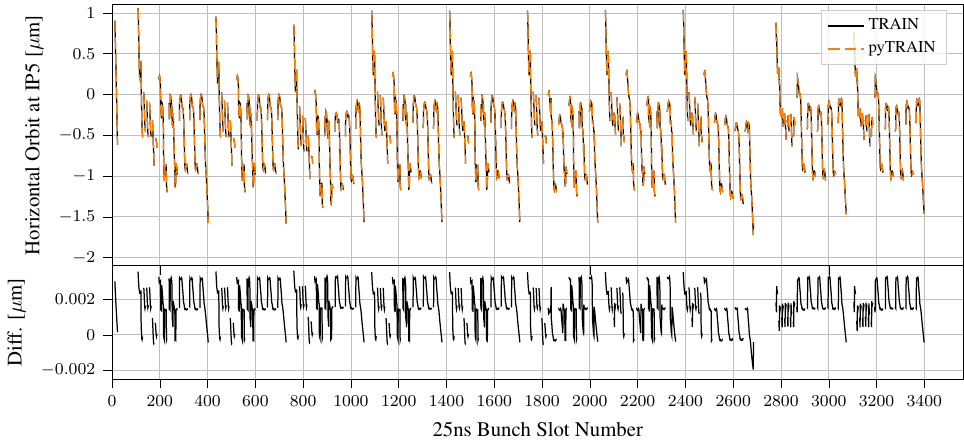}\\[0.5cm]
\includegraphics{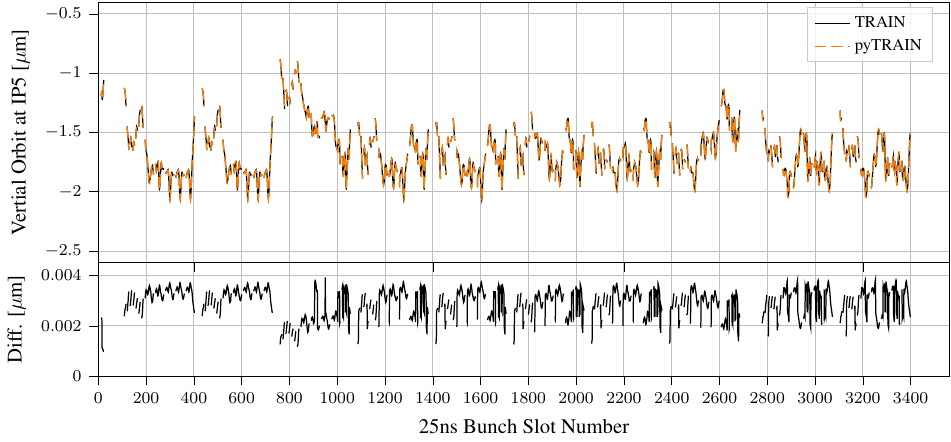}
\caption{Closed orbits of beam 1 at IP5 in a typical 2023 LHC physics scenario, predicted by the classic TRAIN and pyTRAIN codes. The results for the other beams, and in other points of the machine, are similar. The differences are at the level of $10^{-9}\ut{m}$.}
\label{ID6077725_fig:bench}
\end{center}
\end{figure}

For validation, the pyTRAIN code has been benchmarked against the classic TRAIN code for typical LHC physics scenarios. Figure~\ref{ID6077725_fig:bench} gives an example of a benchmark run for a bunch filling pattern and machine configuration used during physics operation in 2023. The differences in closed orbits are at the level of the convergence limit for the iterative solver ($10^{-9}\ut{m}$). During this benchmarking, a problem was identified in the classic TRAIN code in the handling of bunch filling patterns which are different for the two beams; this case is treated correctly in the pyTRAIN code.

\subsection*{Beam Separations from Luminosity Scans}
Beam separation scans in the two high-luminosity experiments ATLAS (IP1) and CMS (IP5) are regularly performed in LHC physics operation. To assess the transverse beam sizes \cite{ID6077725_bib:prablumiscan} and the stability of the experiments luminosity calibration using the Van-der-Meer method \cite{ID6077725_bib:vdm, ID6077725_bib:cmslumicalib}, the two beams are separated in steps at the scanned IP, the change in the bunch-by-bunch luminosity signal from the experiment is recorded. For each bunch pair colliding at the IP, a Gaussian distribution is fitted to the measured luminosity with respect to the introduced separation. From these fits, apart from the beam overlap width, the bunch-by-bunch closed orbit separation can be derived. This is equivalent to the difference in closed orbits of the two bunches encountering at the IP \cite{ID6077725_bib:prablumiscan}.

\begin{figure}[h]
\begin{center}
\includegraphics{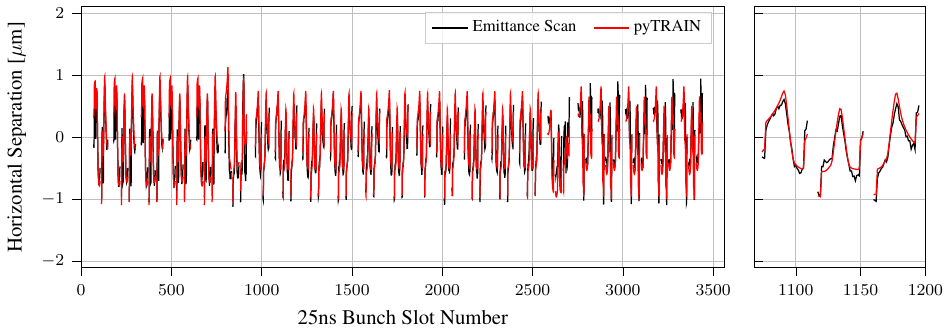}\\[0.5cm]
\includegraphics{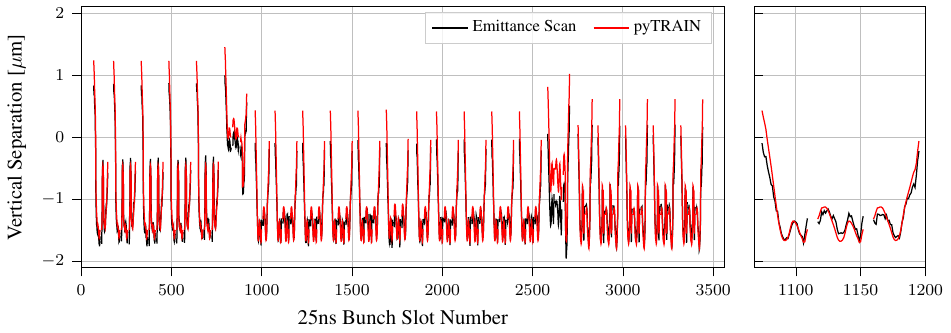}
\caption{Parasitic separation of the colliding bunch pairs at LHC IP5, introduced by long-range beam-beam effects. The right panels show a magnification of one bunch train structure. LHC proton physics fill 10066 (August 2024), at $\beta^*=1.2\ut{m}$ with $1.6\cdot10^{11}$ protons per bunch.}
\label{ID6077725_fig:emitscan}
\end{center}
\end{figure}

This measured closed orbit separation can then be compared to the prediction given by the pyTRAIN code. An example from an LHC 2024 physics fill is given in figure~\ref{ID6077725_fig:emitscan}. All structures predicted by the simulation are observed in the measurement. The observed differences are primarily due to a non perfect reproduction of beam separation in the simulation for the two LHC experiments levelled by beam offsets using a luminosity feedback loop (IP2 and IP8).

\subsection*{Luminous Centroid Positions at the Collision Points}
During collisions, the LHC experiments reconstruct the primary vertices of each recorded interaction using their inner detector (tracker) data. The spatial distribution of the primary vertices gives the size and position (``centroid'') of the luminous region. The position of the luminous region is determined by the average closed orbit offset of the two beams relative to the detector reference system \cite{ID6077725_bib:atlbeamspotrun1, ID6077725_bib:atlbeamspotrun3}.

\begin{figure}[h]
\begin{center}
\includegraphics{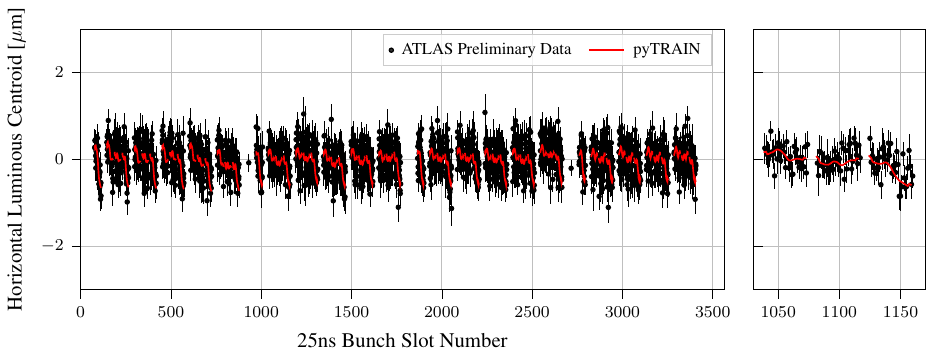}\\[0.5cm]
\includegraphics{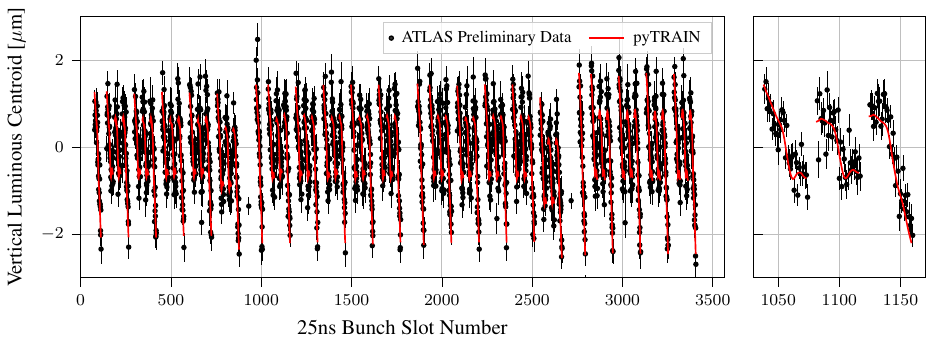}
\caption{Transverse luminous region position (centroid) of the bunch pairs colliding at LHC IP1, as predicted by pyTRAIN and measured by the ATLAS experiment. The right panels show a magnification of one bunch train structure. LHC calibration transfer fill 9635 (May 2024), $\beta^*=30\ut{cm}$ with $1.3\cdot10^{11}$ protons per bunch. The preliminary reconstructed luminous region data is provided courtesy of the ATLAS collaboration \cite{ID6077725_bib:klauswitoldcatrincomm}.}
\label{ID6077725_fig:atlbeamspot}
\end{center}
\end{figure}

In regular proton physics data taking, only a small fraction of the detector read-out capacity is used for reconstruction of the luminous region, which would not allow reconstructing the centroid bunch-by-bunch with a statistical uncertainty low enough to resolve beam-beam orbit effects. However, during special calibration runs, the ATLAS experiment (IP1) has recorded high-rate luminous region data \cite{ID6077725_bib:klauswitoldcatrincomm} which is compared to the beam-beam bunch orbits predicted by pyTRAIN in figure~\ref{ID6077725_fig:atlbeamspot}.

\subsection*{Beam Positions from Wire Scans}
The Wire Scanners, installed in LHC point 4, provide the reference beam size measurement at the LHC \cite{ID6077725_bib:wirescanner}. The wire scan data provides bunch-by-bunch beam sizes and offsets for both beams and planes. While the wire scanners can not be used on a full LHC physics beam at top energy (intensity limit due to beam loss and wire damage thresholds), wire scans are regularly performed of the first bunch trains injected into the LHC.

During injection, the two beams are separated at all interaction points with closed orbit bumps of several millimetres. However, the bunches still experience beam-beam long-range encounters in the vicinity of the IPs. This leads to closed-orbit distortions which can be observed using the wire scanners and predicted by pyTRAIN.

\begin{figure}[h]
\begin{center}
\includegraphics{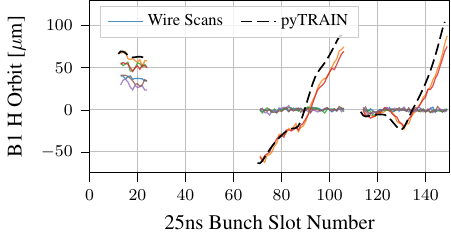}~~~\includegraphics{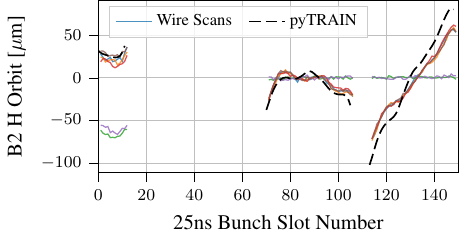}\\[0.5cm]
\includegraphics{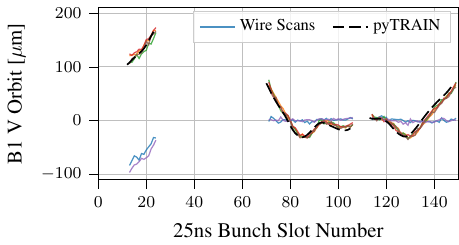}~~~\includegraphics{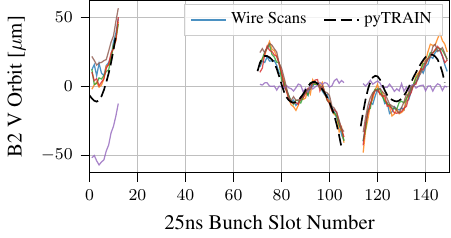}
\caption{Beam offsets measured at the LHC wire scanners during the filling periods in August 2024, and predicted offsets from pyTRAIN considering the long-range beam-beam encounters.}
\label{ID6077725_fig:wirescans}
\end{center}
\end{figure}

Figure~\ref{ID6077725_fig:wirescans} shows bunch offsets measured by the wire scanners in comparison to the pyTRAIN orbit prediction at the wire scanner location. It is worth noting that several wire scans show almost no closed orbit distortion; these measurements were taken when only the first 12 bunches, but no longer bunch trains, were circulating in one of the two beams of the machine.

\section*{Conclusions}
The pyTRAIN code is a new, flexible and robust implementation of the TRAIN concept in python using well established libraries for scientific computing. It simulates self-consistent closed orbits for all bunches in a collider with any number of beam-beam encounter points using a Soft-Gaussian iterative approach. Between the beam-beam encounter points, second-order maps generated by an external code such as MADX or Xsuite are used to describe the machine elements. Once closed-orbit problem for all bunches is solved, the resulting second-order one-turn maps for each bunch can be further analyzed to find tunes, chromaticities, and optical functions for each bunch.

The pyTRAIN code has been successfully benchmarked both against the classic TRAIN code, as well as against measurements taken during LHC physics operation from different sources: beam separation scans, luminous regions measured by the experiments, and wire scans. Results show a very good agreement of the pyTRAIN predictions to the measurements taken at the LHC in different conditions and with various bunch filling patterns.

In the future, a tighter integration with the Xsuite code can be envisaged. While Xsuite can already be used to generate the second-order maps for pyTRAIN, future work could allow for a bi-directional integration, re-using the self-consistent bunch-by-bunch orbits from pyTRAIN as a starting point for tracking simulations using Xsuite.

\end{document}